\documentstyle[12pt]{article}

\input psfig
\begin{document}
\begin{titlepage}
\hfill  UCTP-114-00

\hfill Belle Note 386

\large
\centerline {\bf  CP Results from Belle~\footnote{Talk presented at Beauty 2000 -
7$^{th}$ International Conference on B-Physics at Hadron Machines, 
September 13-18, 2000, Sea of Galilee, Israel}}
\normalsize
 
\vskip 2.0cm
\centerline {Kay Kinoshita~\footnote{kayk@physics.uc.edu}}
\centerline {\it University of Cincinnati, Cincinnati, Ohio, USA}
\vskip 1.0cm
\centerline {For the BELLE Collaboration} 
\vskip 4.0cm
 
\centerline {\bf Abstract}
\vskip 1.0cm
The Belle detector at KEKB has collected 6.2~fb$^{-1}$ of data at the 
$\Upsilon$(4S) resonance.
Presented here are preliminary measurements, from this data sample, 
of sin$2\phi_1$ and other quantities relating to CP.
\vskip 2.0cm
\vfill
\end{titlepage}

\newpage

\begin{figure}[h]
{\centerline
{\psfig{figure=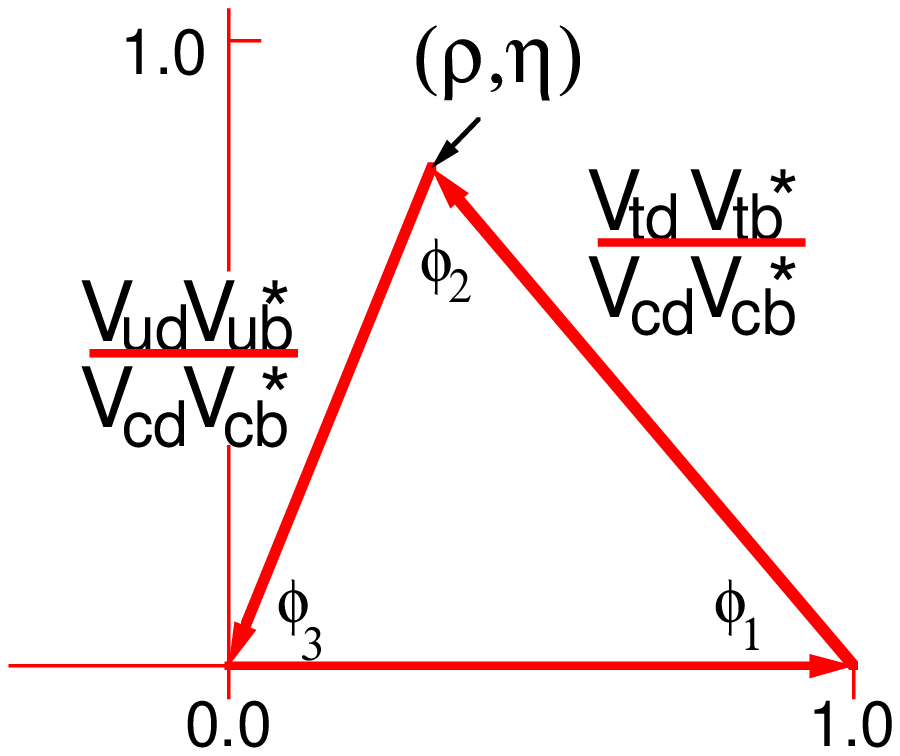,width=5.cm}}
}
\caption{A unitarity triangle.}
\label{fig:triangle}
\end{figure}

In the Standard Model, CP nonconservation occurs naturally in the 
charged current weak interactions of quarks and is manifested by the 
irreducibly complex character of the unitary $3\times 3$ matrix known 
as the CKM matrix.
Essential to the model is the unitarity of the matrix, which reduces 
the nine complex couplings to just four real parameters, often 
represented explicitly in the Wolfenstein parametrization:
\begin{eqnarray}
{\pmatrix{V_{ud} & V_{us} & V_{ub}\cr V_{cd} & V_{cs} & V_{cb}\cr V_{td} & V_{ts} & V_{tb}}} = {\pmatrix{1-\lambda^2/2 & \lambda & \lambda^3A(\rho-i\eta)\cr 
-\lambda & 1-\lambda^2/2 & \lambda^2A\cr 
\lambda^3A(1-\rho-i\eta) & -\lambda^2A & 1}}.
\label{eq:cp}
\end{eqnarray}
A condition of unitarity, that the scalar product of any column with 
the complex conjugate of any other be zero, may be applied to the 
first and third columns to obtain
\begin{eqnarray*}
{V_{td}V_{tb}^*\over V_{cd}V_{cb}^*}+1+{V_{ud}V_{ub}^*\over V_{cd}V_{cb}^*}=0.
\end{eqnarray*}
This may be represented as a closed triangle in the complex plane
with a base of unit length on the real axis and apex coordinates 
($\rho,\eta$) (Fig.~\ref{fig:triangle}).
The angles of this ``unitarity triangle'' may be probed directly 
through CP asymmetries in $B$ meson decay.

The principal goal of the B factory experiments is to test the 
validity of the CKM picture through the observation and measurement 
of these asymmetries.
Among the anticipated measurements, that of sin$2\phi_1$ is 
considered to be the most statistically straightforward.
A nonzero value of $\phi_1$ would be manifested as an ``indirect'' CP 
asymmetry, a difference between the decays of $B^0$ and $\bar B^0$ to 
a common final state of definite CP symmetry.
The rate to final states $f_{CP}$ that are two-body modes with a 
charmonium and neutral K or $\pi^0$  is given by
\begin{eqnarray*}
{dN\over dt}(B\rightarrow f_{CP})={1\over 2}\Gamma e^{-\Gamma\Delta t}
[1+\eta_b\eta_{CP}{\rm sin}2\phi_1{\rm sin}(\Delta m\Delta t)],
\end{eqnarray*}
where $\eta_b=+1(-1)$ for a $B^0(\bar B^0)$,
$\eta_{CP}=+1(-1)$ if CP is even (odd), and $\Delta t$ is the time 
interval from creation to the CP eigenstate decay.
The asymmetry displays an oscillation as a function of the time 
between creation and decay but vanishes upon integration over time.
``Direct'' CP violation results in an asymmetry in integrated rate 
between a decay channel and its charge conjugate and is predicted for 
several modes which are sensitive to other CKM angles: B$\rightarrow 
D^0 K^-$ \{$D^0 \rightarrow K^+ K^-$\} ($\phi_3$), 
B$\rightarrow$K$\pi$, $\pi \pi$, KK ($\phi_2$,$\phi_3$).
Most of these modes are CKM-suppressed so that they are not only 
rare, they must contend with specific backgrounds from their 
CKM-favored counterparts.
Many of them have yet to be observed.

The Belle experiment at the KEKB ring is designed to measure both 
types of asymmetry.
At KEKB, 8.0 GeV electrons and 3.5 GeV positrons annihilate to 
produce the $\Upsilon$(4S) resonance in motion with 
$\beta\gamma=0.425$, and the experiment relies on this motion to 
measure the time-dependent asymmetry.
The $\Upsilon$(4S) is CP-odd and decays exclusively to $B\bar B$ pairs.
The CP symmetry is conserved until the instant of the first $B$ 
decay, which may thus be considered the instant of creation for the 
remaining $B$.
The initial flavor of this $B$ may be known through inspection of the 
first decay, a process known as ``flavor tagging.''
The second decay is reconstructed in CP-eigenstate channels, and is 
termed the ``CP tag.''
Because the two $B$'s move slowly in the $\Upsilon$(4S) 
center-of-mass (CM) frame ($\beta\gamma=0.06$), their lab velocities 
are nearly equal, so that the displacement between their decay 
vertices is proportional to $\Delta t$.

For all CP measurements the experiment will need a large sample of events.
KEKB has achieved a luminosity of $2.0\times 
10^{33}$cm$^{-2}$s$^{-1}$, and hopes to reach the design luminosity 
of  $1\times 10^{34}$cm$^{-2}$s$^{-1}$ in the next year.
Between June 1999 and July 2000 the Belle experiment logged 
6.2~fb$^{-1}$ of integrated luminosity on the $\Upsilon$(4S) 
resonance, plus 0.6~fb$^{-1}$ at an energy just below the resonance.
The preliminary estimate of the number of $B\bar B$ events in this 
sample is $6.34\times 10^6$.

To measure sin2$\phi_1$, we first reconstruct CP tags in two-body 
decays to charmonium and $K_S$, $K_L$, or $\pi^0$.
Charmonia are reconstructed in the modes $J/\psi \rightarrow 
\ell^+\ell^-$, $\psi(2S) \rightarrow \ell^+\ell^-$, $\psi(2S) 
\rightarrow J/\psi \pi^+\pi^-$, and $\chi_c\rightarrow J/\psi \gamma$.
Each event is then passed through a flavor tagging algorithm to 
determine the flavor of the other $B$ decay.
In each tagged event the two vertices are measured to obtain $\Delta t$.
An unbinned maximum likelihood fit to sin2$\phi_1$ is then performed 
on the distribution in $\Delta t$,
accounting for the CP and flavor of each event.
To make this measurement one must account for the backgrounds to the 
CP tag, the fraction of incorrect flavor tags, and various sources of 
uncertainty on the measurement of $\Delta t$.

\begin{figure}[h]
{\centerline
{\psfig{figure=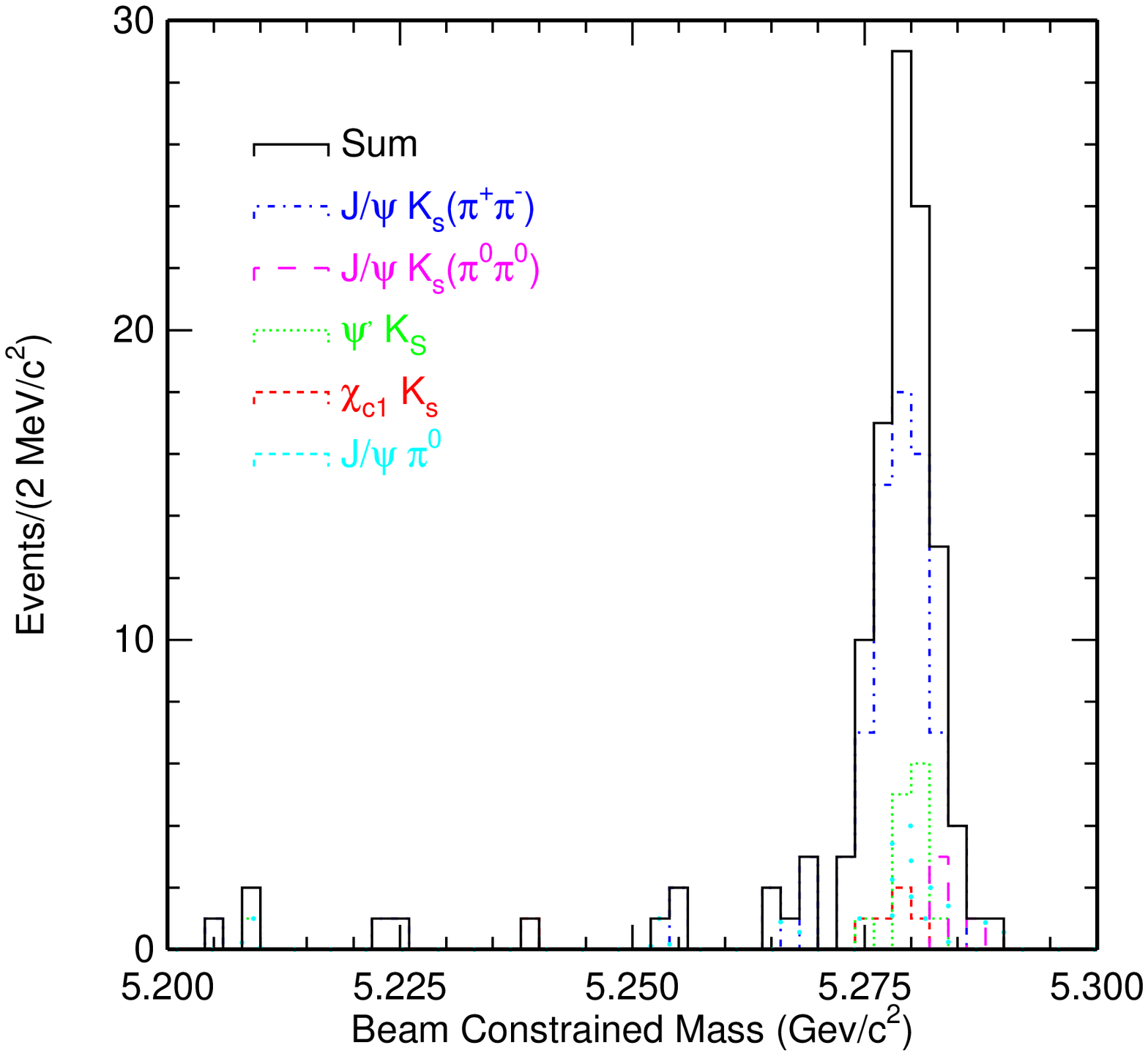,width=6.cm}\psfig{figure=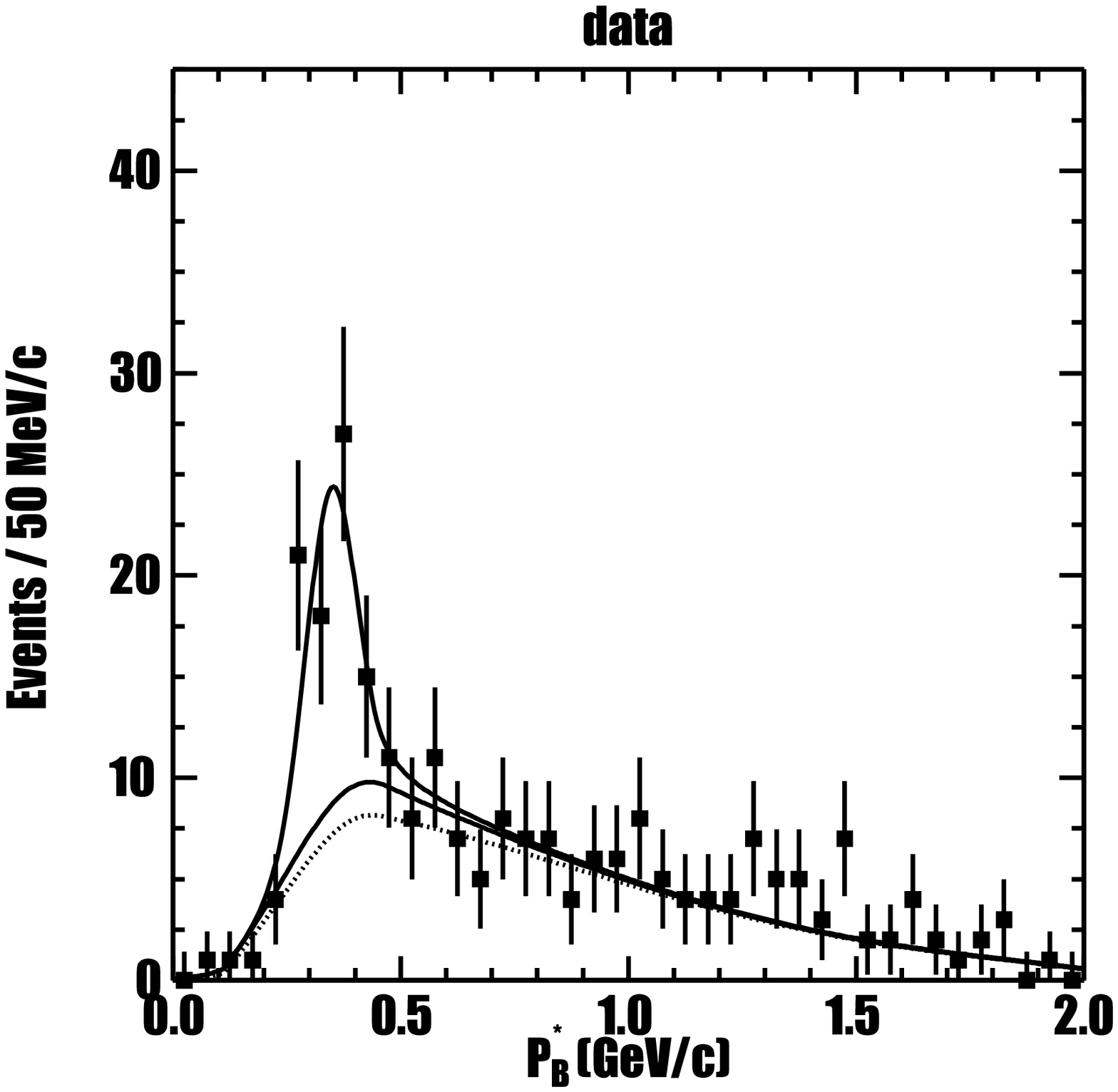,width=6.cm}}
}
\caption{(Left) Distribution in $M_B$ of all fully reconstructed $B$ 
candidates in charmonium plus $K_S$/$\pi^0$ final states.
(Right) Distribution in $p_B^*$ of $B$ candidates to charmonium plus 
$K_L$ final states, with fits to signal plus background.}
\label{fig:psik}
\end{figure}

For fully reconstructed modes, criteria for charged particle identification and invariant
mass are unrestrictive, designed to maximize  efficiency while achieving good
signal/background in the final sample.
The signal is observed through the distributions of candidates in $\Delta E$ and $M_B$; 
 $\Delta E$, the difference between the candidate's reconstructed energy and the CM beam
energy, centers at zero for signal with a width that is dominated by detector resolution
(10-50 MeV), and $M_B=\sqrt{E_{beam}^{*2}-p^{*2}_{cand}}$, known as the beam-constrained
mass, has a width determined by detector momentum resolution and accelerator beam energy
spread.  
Figure~\ref{fig:psik} shows the distribution in $M_B$ of all fully reconstructed CP tag
mode candidates that pass mode-dependent requirements on $\Delta E$.

To reconstruct $J/\psi K_L$ decays, we impose tighter constraints on 
the $J/\psi$ candidates and require that the CM momentum fall within 
the range 1.42-2.00~GeV/c.
The observed CM momentum of the $J/\psi$ determines the average 
4-momentum of the recoiling $K_L$.
A $J/\psi K_L$ candidate is constructed if a $K_L$ candidate is found 
within 45$^\circ$ of the expected direction in the lab frame.
The CM momentum of the candidate, $p^*_{cand}$, may then be evaluated 
by assuming that its mass is that of the $B$.
The distribution in $p^*_{cand}$ is fitted to a sum of expected 
distributions from signal and backgrounds, as shown in 
Fig.~\ref{fig:psik}.
The shapes are estimated by Monte Carlo simulation.
True $J/\psi K_L$ decays accumulate near the known $B$ momentum, 
around 0.3~GeV/c.
Most of the background can be accounted for from known higher 
multiplicity $B$ decays that include $\psi$ and $K_L$ in the final 
state, such as $B\rightarrow\psi K^*$.

For each event containing a CP tag, the remainder of the event is 
examined to determine the original flavor of the other $B$.
The highest efficiencies for flavor tagging have been obtained 
through identification of single tracks, where the sign of the charge 
is correlated with flavor; high-momentum ($p^*>1.1$~GeV/c) electrons 
or muons from $b\rightarrow c\ell^-\bar\nu$, $K^-$ from $b\rightarrow 
cX \{c\rightarrow sY$\}, lower momentum leptons from $b\rightarrow cX 
\{c\rightarrow \ell^+Y$\}, and soft pions from $b\rightarrow D^{*+}X 
\{D^{*+}\rightarrow D^0\pi^+$\}.
Only the first two of these methods are used for the results presented here.
The numbers of reconstructed decays without and with flavor tags are 
summarized in Table~\ref{tab:CPtags}.
\begin{table}[htb]
\centering
\begin{tabular}{ l c c c c} \hline
  & Decay Mode & \# cands & Background & \# tagged \\ \hline
  CP=$-$1 & $J/\psi K_S$, $K_S \rightarrow \pi^+\pi^-$ & 70 & $3.4\pm 
1.0$ & 40 \\
  & $J/\psi K_S$, $K_S \rightarrow \pi^0\pi^0$ & 4 & $0.3\pm 0.1$ &4 \\
  & $\psi(2S) K_S$, $\psi(2S) \rightarrow \ell^+\ell^-$&5 & $0.2\pm 0.1 $ &2 \\
  & $\psi(2S) K_S$, $\psi(2S) \rightarrow J/\psi\pi^+\pi^-$&8 & 
$0.6\pm 0.3 $ &3 \\
  & $\chi_{c1} K_S$ & 5 & $0.8\pm 0.4$ &3 \\
\hline
  CP=$+$1 & $J/\psi K_L$ & 102 & $47.6\pm 4.8$ & 42 \\
  & $J/\psi \pi^0$ & 10 & $0.6\pm 0.3$ & 4 \\
  \hline
  & Total& 204 &   & 98\\
  \hline
\end{tabular}
\caption{Tagging summary.  CP modes, number of candidates, estimated 
background, number of candidates with flavor tags.}
\label{tab:CPtags}
\end{table}

To determine the decay time difference, we measure the decay vertices of both $B$ decays
in each tagged event.
For the CP decay, only the $J/\psi \rightarrow \ell^+\ell^-$ candidate is used to
reconstruct the vertex.
The resolution of this vertex in the beam $(z)$ direction is around 40$\mu$m.
The vertex of the flavor-tagged $B$ is estimated by examining the tracks not associated
with the CP decay, excluding tracks from identified $K_S$.
The tracks are fitted to a single vertex, using
an iterative procedure to exclude tracks that fit poorly.
The resolution of this measurement is limited not only by detector resolution but also by
the event itself, which may contain secondary charm particles and thus not have a unique
vertex.
The resolution is found via Monte Carlo simulation to be about 85$\mu$m.
The net resolution on the difference, $\Delta z$, is 100$\mu$m.

The decay time difference  $\Delta t$ is calculated as the measured 
difference in the reconstructed z-coordinates, $\Delta z$, divided by 
$\beta\gamma c$ ($\beta\gamma=0.425$).
The fitting function takes into account the root distribution of the 
signal (an analytic function), the fraction of incorrect flavor tags 
(assumed to be constant), background to reconstructed CP decays (both 
correctly and incorrectly tagged), and the resolution of $\Delta t$, 
parametrized and assigned event-by-event.

\begin{figure}
{\centerline
{\psfig{figure=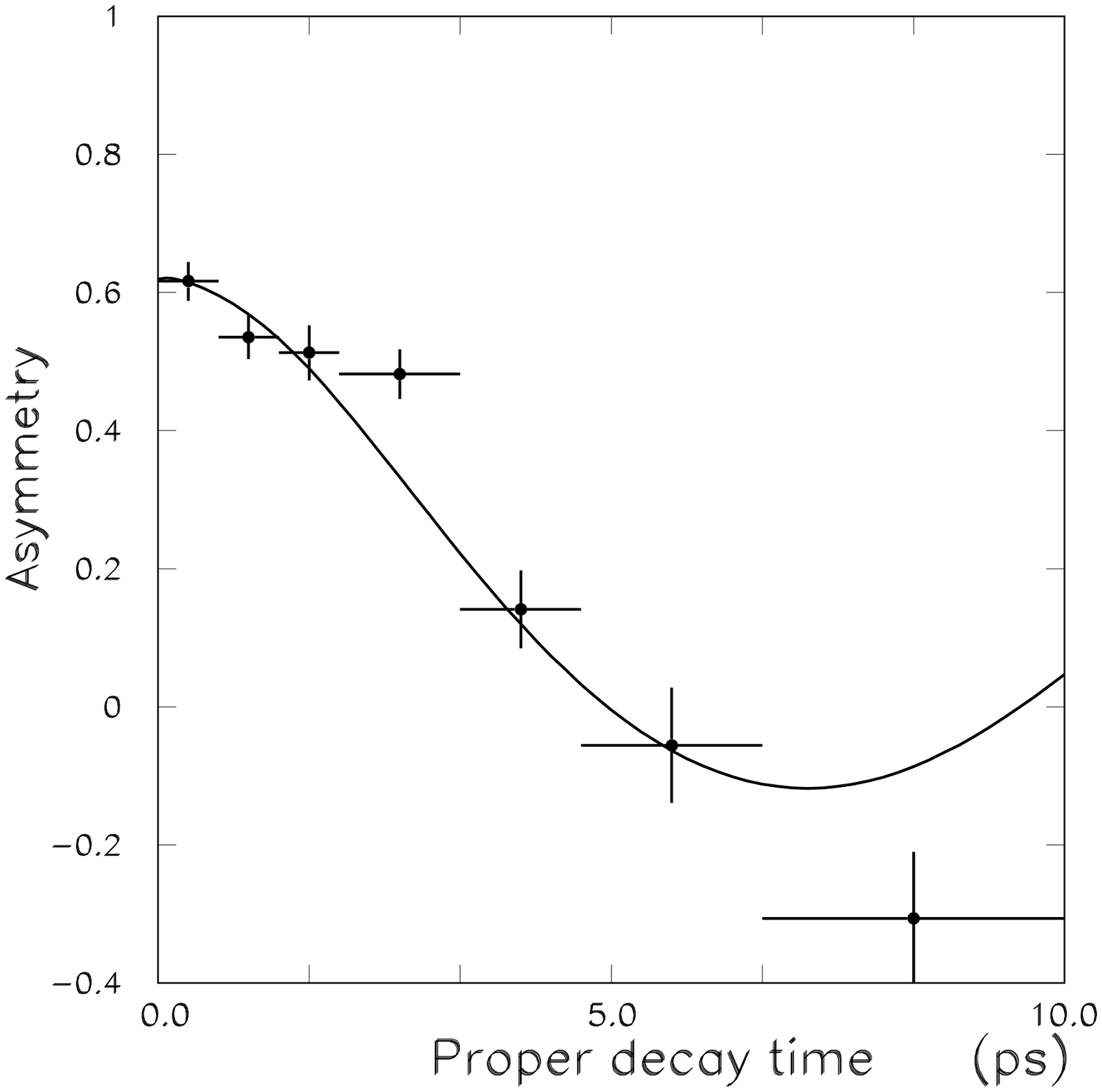,width=6.5cm}
\psfig{figure=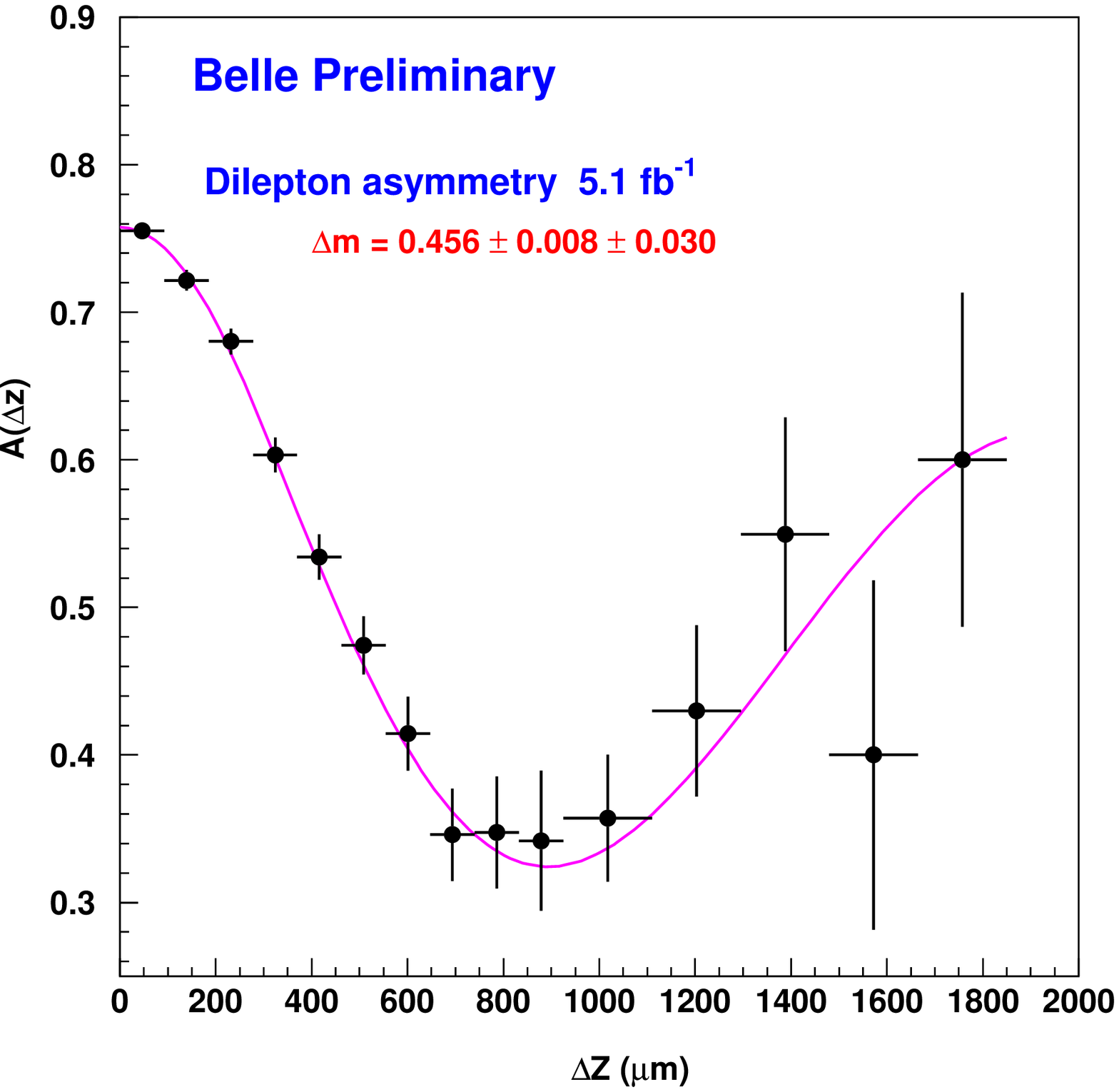,width=6.5cm}}
}
\caption{Asymmetry of opposite- and same-flavor events.  
(Left) Event flavors were tagged using decays $B\rightarrow D^{(*)}\ell\nu$ on one side
and flavor tagging on the other, as described in the text.  Plotted as a function of
proper decay time.
(Right) Event flavors were tagged using dileptons.  Plotted as a function of $\Delta z$.} 
\label{fig:asym_mixing} 
\end{figure}

It is important that the rate and distribution of incorrect flavor tags be evaluated
accurately.
The wrong tag fraction is measured using the same fitting method and data sample as the CP
fit, but where a flavor-specific decay replaces the CP decay.  
Two flavor-specific modes, $B\rightarrow D^{*-}\ell^+\nu$ and $B\rightarrow
D^{-}\ell^+\nu$ (and the charge conjugate modes),  are reconstructed.
The events are separated into same- and opposite-flavor samples.
The asymmetry in the $\Delta t$ distributions contains an oscillation due to mixing:
$A_{mix}={N_{opp}(\Delta t)-N_{same}(\Delta t)\over 
        N_{opp}(\Delta t)+N_{same}(\Delta t)}
=(1-2w){\rm cos}(\Delta m_d\Delta t)$.
The parameters of the fit include the wrong tag fractions for neutral  ($w$) and charged 
($w^+$) $B$ events, the mixing parameter $\Delta m_d$, and the detector response function.
A fit for $\Delta m_d$ yields a clear mixing oscillation (Figure~\ref{fig:asym_mixing},
left), with
$\Delta m_d=0.49\pm 0.026$~ps$^{-1}$.

Our measurement of sin$2\phi_1$ includes contributions from a variety 
of CP modes and flavor tag methods with varying efficiencies, 
backgrounds, and wrong tag rates.
The contribution of each flavor tag is quantified by the ``effective 
tagging efficiency'',
$\epsilon_{eff}=(1-2w)^2\epsilon_{tag}$.
To obtain the wrong tag fractions, each tag is evaluated separately.
Efficiencies and effective efficiencies are summarized in 
Table~\ref{tab:efftags}.

\begin{table}[htb]
\centering
\begin{tabular}{c c c c} \hline
Tag & $\epsilon_{tag}$(\%)  & $w$(\%) & $\epsilon_{eff}$(\%) \\ \hline
high-$p^*$ lepton & $14.2\pm 2.1$ & $7.1\pm 4.5$ & $10.5\pm 2.7 $ \\
Kaon & $27.9\pm 4.2$ & $19.9\pm 7.0$ & $10.1\pm 4.9$ \\
medium-$p^*$ lepton & $2.9\pm 1.5$ & $29.2\pm 15.0$ & $0.5$ \\
soft $\pi$ & $7.0\pm 3.5$ & $34.1\pm 15.0$ & $0.7 $
\\ \hline
Total & 52.0 & & 21.2
\\ \hline
\end{tabular}
\caption{Tagging efficiency ($\epsilon_{tag}$), wrong tag fraction 
($w$), and effective tagging efficiency ($\epsilon_{eff}$) for the 
flavor tagging methods.}
\label{tab:efftags}
\end{table}

The response function must account accurately for detector resolution 
and physics effects in data.
Based on Monte Carlo studies, we use a response function that is a 
parametrized sum of two Gaussians and takes into account the detector 
resolution, the effect of mismeasured tracks, biases from finite 
lifetimes of charm particles, and the approximation $\Delta t=\Delta 
z/\beta\gamma c$.
The parameters are determined event-by-event, based on the quality of 
track fitting.

The response function may be tested by measuring known lifetimes.
The charged and neutral $B$ lifetimes were measured with the 
technique used for the CP asymmetry: $B$ reconstruction on one side 
and a flavor tag on the other.
The $B$'s were reconstructed in exclusive hadronic and semileptonic modes.
We obtain
$\tau_{B^0}=1.50\pm 0.05\pm 0.07$~ps and $\tau_{B^-}=1.70\pm 0.06\pm 
0.11$~ps, which are consistent with the most recent world 
averages\cite{pdg2000}, $1.548\pm 0.032$~ps  and $1.653\pm 0.028$~ps.
We also measure the neutral B mass difference $\Delta m_d=0.456\pm 
0.008\pm 0.030$~ps$^{-1}$ in a dilepton analysis\cite{mixing}.
Figure~\ref{fig:asym_mixing}(right) shows a clear mixing oscillation 
in the net opposite/same-sign dilepton asymmetry as a function of 
$\Delta z$.

\begin{figure}
{\centerline
{\psfig{figure=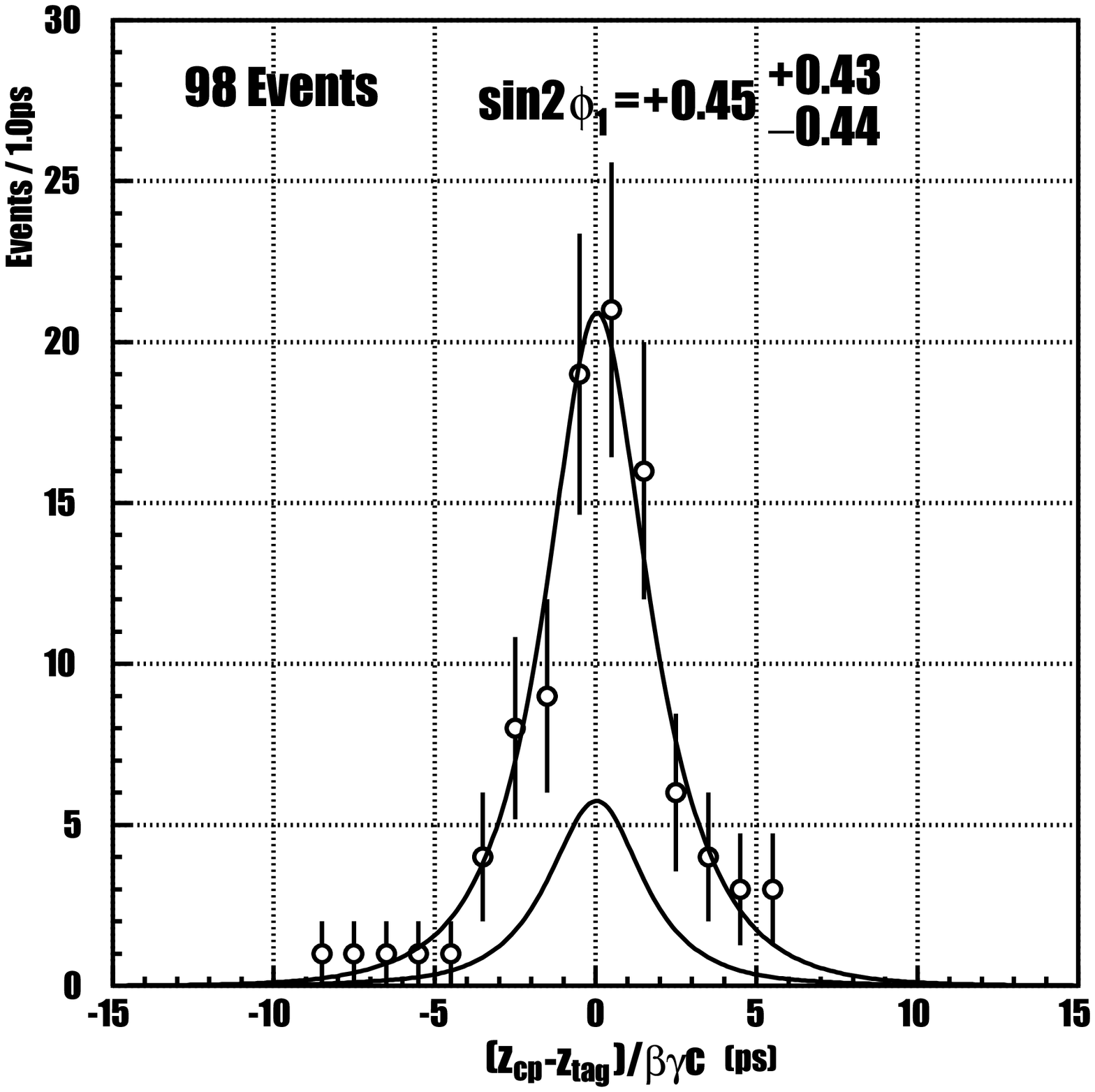,width=6.cm}
\psfig{figure=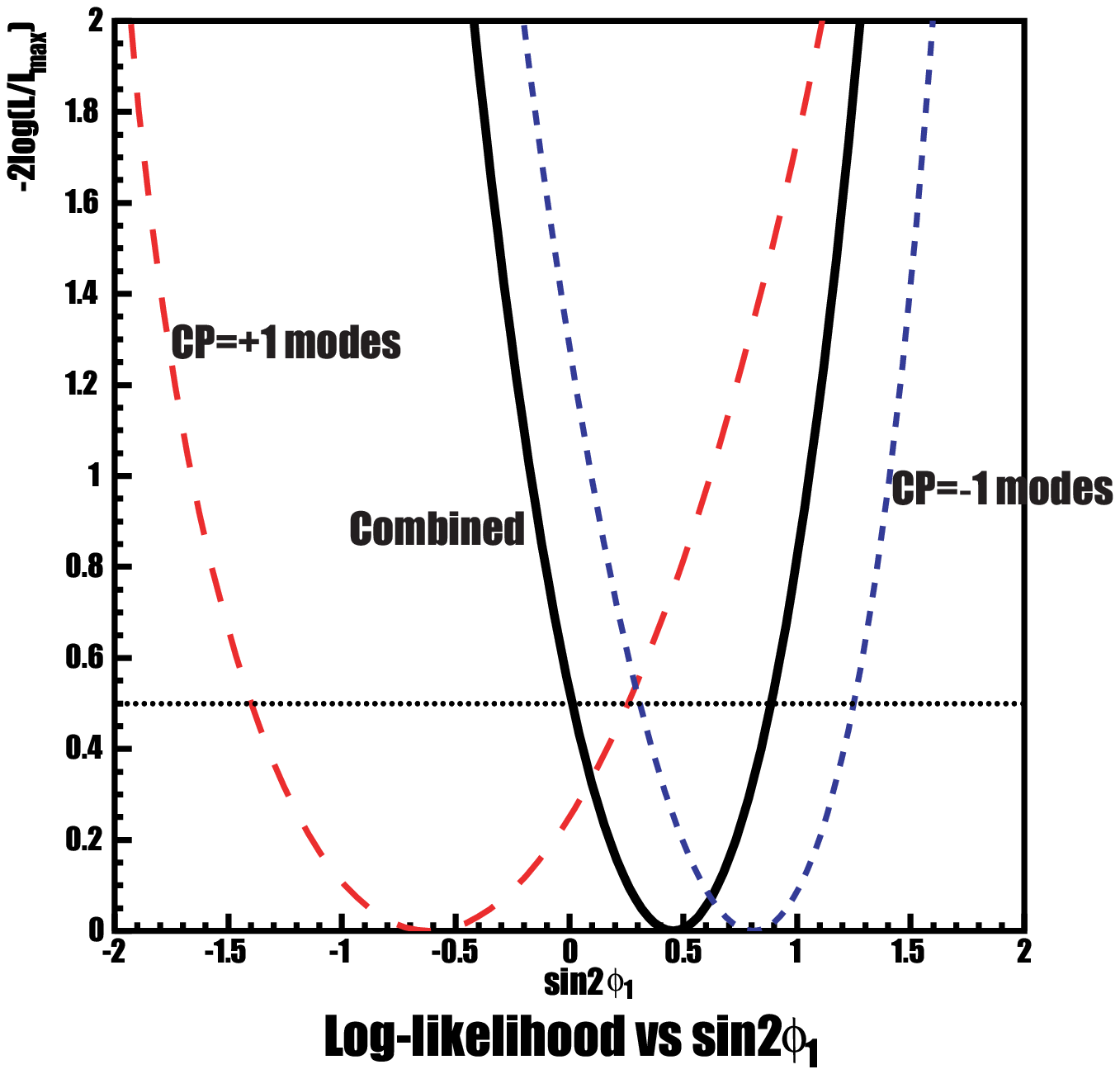,width=6.5cm}}
}
\caption{(Left) Distribution in $\eta_b\eta_{CP}\Delta t$ (see Eq.~\protect\ref{eq:cp}) of 98
events with reconstructed CP decay and flavor tag.
Upper curve shows result of fit for sin$2\phi_1$.  Lower curve shows contribution of
background to this fit.
(Right) Log likelihood distribution for CP=$-1$, CP=$+1$, and combined tags.
}
\label{fig:sinphi1}
\end{figure}

Figure~\ref{fig:sinphi1} shows the distribution in data and the result of the fit for
sin$\phi_1$ with all tags combined, as well as the log-likelihood distributions for the
CP$=-1$, CP$=+1$, and combined tags.
We find\cite{sinphi1}
\begin{eqnarray*}
{\rm sin}2\phi_1=0.45^{+0.43+0.07}_{-0.44-0.09}.
\end{eqnarray*}
This is not yet sufficiently precise to contribute significantly to 
current constraints on the CKM matrix.
Continuing work includes the addition of CP modes, development of 
flavor tags, and accumulation of more data.

The mode $B^0\rightarrow J/\psi K_1^0(1270)\{K_1(1270)\rightarrow K_S\rho^0\}$, is a CP
eigenstate.
We have made the first observations of decays $B\rightarrow J/\psi K_1(1270)$ using
$J/\psi$ reconstructed in the dilepton modes and $K_1(1270)\rightarrow K\rho$
reconstructed in $K^+\pi^+\pi^-$, $K^+\pi^-\pi^0$, and $K^0\pi^+\pi^-$\cite{psik1}.
The $\Delta E$ distributions for all analyzed modes are combined and fitted to obtain the
net signal, $45.0^{+8.7}_{-8.3}$ candidates in 5.3~fb$^{-1}$ of data.
The branching fractions are obtained by first taking the ratios of the observed modes with 
$B^+\rightarrow J/\psi K^+$, then multiplying by the  branching fraction ${\cal
B}(B^+\rightarrow J/\psi K^+)=(9.9\pm 1.0)\times 10^{-4}$\cite{pdg1998}.
We obtain
\begin{eqnarray*}
& {\cal B}(B^0\rightarrow J/\psi K_1^0(1270))=(1.4\pm 0.4\pm 
0.4)\times 10^{-3}\\
{\rm and} & {\cal B}(B^+\rightarrow J/\psi K_1^+(1270))=(1.5\pm 
0.4\pm 0.4)\times 10^{-3}.
\end{eqnarray*}

The CP eigenvalue of the mode $B^0\rightarrow J/\psi K^{*0} 
(K^{*0}\rightarrow K_S\pi^0)$ depends on the helicity of the $K^*$.
This mode could be useful for CP studies if its polarization favors 
one eigenstate.
We reconstruct $B\rightarrow J/\psi K^{*}$ through $J/\psi 
\rightarrow \ell^+ \ell^-$ and K*$\rightarrow  K^+ \pi^-$,  $K_S 
\pi^+$, $K^+ \pi^0$ and
find 176 candidates in 5.1~fb$^{-1}$ of data.
Their distributions in helicity and transversity are fitted to 
obtain\cite{psikst}
\begin{eqnarray*}
&\Gamma_L/\Gamma=0.52\pm 0.06 \pm 0.04 \\
{\rm and }&
|A_\perp|^2=0.27\pm0.11\pm0.05.
\end{eqnarray*}
These values indicate that $CP$-even states dominate in 
$B^0\rightarrow J/\psi K^{*0}$.

\begin{figure}
{\centerline
{\psfig{figure=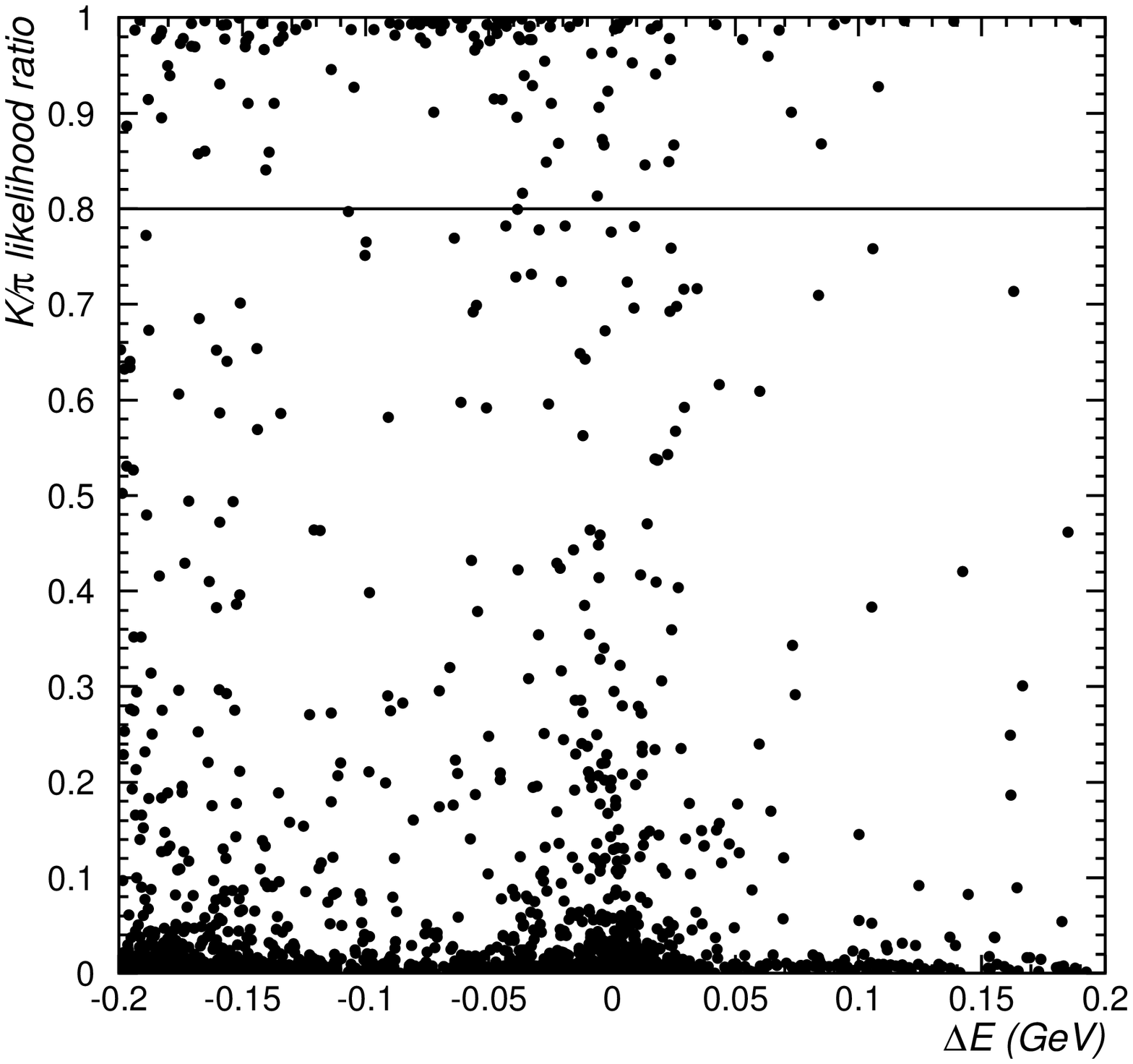,width=6.cm}}
}
\caption{Distribution of $\bar B^0\rightarrow D^{*+}\pi^-$ candidates in $\Delta E$ and
KID.  
Candidates cluster at $\Delta E \approx -0.05$, as expected for decays $\bar
B^0\rightarrow D^{*+}K^-$, where KID$\sim 1$.}
\label{fig:DK}
\end{figure}

The decay rates for $B\rightarrow  D^{(*)}$K and $B\rightarrow  D^0 K^-$ \{$D^0
\rightarrow f_{CP}$\} are sensitive to $\phi_3$.
Good separation of $K$ from $\pi$ is required to distinguish this decay from the much
larger signal of the kinematically similar Cabibbo-favored decay $B \rightarrow
D^{(*)}\pi$.
Based on information from the aerogel \v{C}erenkov counter, $dE/dx$ in the drift chamber,
and time-of-flight, our Kaon identification parameter (KID) is a likelihood ratio that
equals 1  (0) for strongly  identified Kaons (pions).
Clear evidence for the DK mode may be seen in the two-dimensional distribution of $\Delta
E$ vs.  KID (Figure~\ref{fig:DK}).
We measure the ratios
${\cal B}(B^- \rightarrow  D^{(*)} K^-)/{\cal B}(B^- \rightarrow  D^{(*)}\pi^-$).
As the reconstructions of these decays are nearly identical, the systematic uncertainties
other than those from identification of the $\pi$ or $K$ cancel in taking the ratio.
We obtain\cite{DK}
\begin{eqnarray*}
{\cal B}(B^- \rightarrow  D^0 K^-)/{\cal B}(B^- \rightarrow 
D^0\pi^-)&=&0.081\pm 0.014\pm 0.011, \\
{\cal B}(B^- \rightarrow D^{*0} K^-)/{\cal B}(B^- \rightarrow D^{*0} 
\pi^-)&=&0.134+0.045\pm 0.015, \\
{\rm and}\  {\cal B}(B^- \rightarrow  D^{*+} K^-)/{\cal B}(B^- 
\rightarrow  D^{*+} \pi^-)&=&0.062\pm 0.030\pm 0.013.
\end{eqnarray*}

We have reconstructed many $K\pi$, $\pi \pi$, and $KK$ final states
\cite{kpi}.
The results are summarized in Table~\ref{tab:kpi}.

\begin{table}[htb]
\centering
\begin{tabular}{c c c c c} \hline
Mode & Yield & $\epsilon$ & ${\cal B} \times 10^5$ & UL$\times 10^5$ \\ \hline
$K^+\pi^-$ & $25.6^{+7.5}_{-6.8}\pm 3.8 $  & $0.28\pm 0.04$ & 
$1.74^{+0.51}_{-0.46}\pm 0.34$ & --  \\
$\pi^+\pi^-$ & $9.3^{+5.7}_{-5.1}\pm 2.0 $  & $0.28\pm 0.04$ & 
$0.63^{+0.39}_{-0.35}\pm 0.16$ & 1.65 \\
$K^+K^-$ & $0.8^{+2.1}_{-0.8} $ & $0.20\pm 0.03$ & -- & 0.6 \\
$K^0\pi^-$ & $5.7^{+3.4}_{-2.7}\pm 0.6 $  & $0.13\pm 0.02$ & 
$1.66^{+0.98}_{-0.78}\pm 0.24$ & 3.4 \\
$K^0K^+$ & $0.0^{+0.5}_{-0.0}$ & $0.11\pm 0.02$ & -- & 0.8 \\
$K^+\pi^0$ & $32.3{^{+9.4}_{-8.4}} {^{+2.4}_{-2.2}} $  & $0.31$ & 
$1.88^{+0.55}_{-0.49}\pm 0.23$ & -- \\
$K^0\pi^0$ & $5.4{^{+5.7}_{-4.4}} {^{+1.0}_{-1.1}}$ & $0.30$ & 
$0.33^{+0.35}_{-0.27}\pm 0.07$ & 1.0 \\
$\pi^+\pi^0$ & $10.8{^{+4.8}_{-4.0}} {^{+0.7}_{-0.5}} $  & 0.19 & 
$2.10^{+0.93}_{-0.78}\pm 0.25$ & -- \\
\hline
\end{tabular}
\caption{$B$ decays to $\pi\pi$, $K\pi$, and $KK$: yield, efficiency ($\epsilon$),
branching fraction (${\cal B}$), and upper limit (UL).} 
\label{tab:kpi}
\end{table}

To summarize, the Belle experiment collected 6.2~fb$^{-1}$ of data at 
the $\Upsilon$(4S) resonance between June 1999 and July 2000.
Preliminary results based on these data include a measurement with
98 tagged events of sin2$\phi_1=0.45^{+0.43+0.07}_{-0.44-0.09}$, the 
first observation of $B\rightarrow\psi K_1$(1270), a measurement of 
$\psi K^*$ polarization, and measurements of other $B$ decay modes 
expected to be used in CP measurements: $D^{(*)}K$,  $K\pi$, $\pi 
\pi$, $KK$.

In the near future we expect to apply additional modes and flavor 
tags toward the measurement of sin2$\phi_1$.
KEKB will resume operations on October 1, 2000 and is expected to run 
with higher currents.

\newpage
\centerline{FIGURES}


\begin{thebibliography}{99}

\bibitem{pdg2000}
D.E. Groom {\it et al}, The European Physical Journal {\bf C15}, 1 (2000).

\bibitem{mixing}
K. Abe {\it et al.}, Belle collaboration, hep-ex/0011090, submitted 
to {\it Physical Review Letters}.

\bibitem{sinphi1}
H. Aihara, hep-ex/0010008 (2000).

\bibitem{pdg1998}
C. Caso {\it et al}, The European Physical Journal {\bf C3}, 1 (1998).

\bibitem{psik1}
Belle Collaboration, ``Observation of $B\rightarrow J/\psi K_1(1270)$,''
KEK preprint 2000-75,
http://bsunsrv1.kek.jp/conferences/ICHEP2000/conf0004.ps.

\bibitem{psikst}
Belle Collaboration, ``Measurement of Polarization of $J/\psi$ in
$B^0\rightarrow J/\psi K^{*0}$ and $B^+\rightarrow J/\psi K^{*+}$ 
Decays,'' KEK preprint 2000-83,\\
http://bsunsrv1.kek.jp/conferences/ICHEP2000/conf0014.ps.

\bibitem{DK}
Belle Collaboration, ``Observation of Cabibbo suppressed 
$B\rightarrow D^{(*)}K$ decays at Belle,''
KEK preprint 2000-80,\\
http://bsunsrv1.kek.jp/conferences/ICHEP2000/conf0010.ps.

\bibitem{kpi}
Belle Collaboration, ``Charmless Hadronic $B$ Meson Decays to Charged 
Particle Final States with Belle,''
KEK preprint 2000-76,
http://bsunsrv1.kek.jp/conferences/ICHEP2000/conf0005.ps;\\
``A
Study
  of Charmless Hadronic B decays to $h\pi^0$ Final States,''\\
KEK preprint 2000-77, \\
http://bsunsrv1.kek.jp/conferences/ICHEP2000/conf0006.ps.

\end{thebibliography}
\end{document}